\documentclass[useAMS,usenatbib,onecolumn]{mnras}
\usepackage{amsmath,bm,amssymb}
\usepackage{dcolumn}
\usepackage[utf8]{inputenc}
\usepackage{graphics,graphicx}
\usepackage{float}
\usepackage{threeparttable}
\usepackage{url}
\usepackage{epstopdf}
\usepackage{soul}

\usepackage{color}

\newcommand{\p}{^\prime}
\newcommand{\pp}{^{\prime\prime}}

\title[ExoMol line lists -- XLI. KOH \& NaOH]{ExoMol line lists -- XLI. High-temperature molecular line lists for the alkali metal hydroxides KOH and NaOH}

\date{\today}

\author[A. Owens et al.]
{A. Owens,\thanks{The corresponding author: alec.owens.13@ucl.ac.uk} J. Tennyson\thanks{The corresponding author: j.tennyson@ucl.ac.uk} and S. N. Yurchenko\thanks{The corresponding author: s.yurchenko@ucl.ac.uk}\vspace*{4mm}\\
Department of Physics and Astronomy, University College London, Gower Street, WC1E 6BT London, UK}

\date{Accepted XXXX. Received XXXX; in original form XXXX}

\pagerange{\pageref{firstpage}--\pageref{lastpage}} \pubyear{2020}

\begin{document}

\label{firstpage}

\maketitle

\begin{abstract}
Potassium hydroxide (KOH) and sodium hydroxide (NaOH) are expected to occur in the atmospheres of hot rocky super-Earth exoplanets but a lack of spectroscopic data is hampering their potential detection. Using robust first-principles methodologies, comprehensive molecular line lists for KOH and NaOH that are applicable for temperatures up to $T=3500$~K are presented. The KOH OYT4 line list covers the 0\,--\,6000~cm$^{-1}$ (wavelengths $\lambda > 1.67$~$\mu$m) range and comprises 38 billion transitions between 7.3 million energy levels with rotational excitation up to $J=255$. The NaOH OYT5 line list covers the 0\,--\,9000~cm$^{-1}$ (wavelengths $\lambda > 1.11$~$\mu$m) range and contains almost 50 billion lines involving 7.9 million molecular states with rotational excitation up to $J=206$. The OYT4 and OYT5 line lists are available from the ExoMol database at \href{http://www.exomol.com}{www.exomol.com} and should greatly aid the study of hot rocky exoplanets.
\end{abstract}


\section{Introduction}

Simple molecules composed of rock-forming elements, such as potassium hydroxide (KOH) and sodium hydroxide (NaOH), are expected to be present in the atmospheres of hot rocky super-Earth exoplanets~\citep{12ScLoFe.exo,jt693}. This class of exoplanets are in close proximity to their host star and tidally-locked, with their dayside exposed to extremely high temperatures, e.g.\ up to 4000~K. The material on the surface of the planet will vaporise to create an atmosphere strongly dependent on the initial planetary composition~\citep{09ScFexx,11MiKaFe}. Probing the spectroscopy of hot rocky super-Earths poses its own unique set of challenges, notably the availability of molecular line lists for relevant molecules composed of rock-forming elements, and the completeness of these line lists at such high temperatures.

The possible detection of KOH and NaOH in exoplanets could help identify terrestrial planets with molten surfaces~\citep{13HaAbGe}, and lead to a better understanding of the relationship between the atmosphere and crust of hot rocky exoplanets at high temperatures ($\gtrsim 3500$~K)~\citep{20HeWoHe}. However, only a small amount of spectroscopic data is available for the alkali metal hydroxides KOH and NaOH, which are linear triatomic molecules in the gas phase. The Cologne Database for Molecular Spectroscopy (CDMS)~\citep{CDMS:2001,CDMS:2005} lists a number of pure rotational transitions in the microwave region for KOH and NaOH as they are considered to be interstellar molecules, despite only the tentative identification of NaOH in the molecular cloud Sagittarius B2~\citep{82HoRhxx.NaOH} and star-forming region Orion-KL~\citep{91Tuxxxx.NaOH}. Aside from the CDMS data, which is based on several laboratory measurements of these molecules~\citep{76PeWiTr.KOH,75KuToDy.KOH,87RaYaGia.KOH,96KaSuHi,96CeOlRi.KOH,73PeTrxx.NaOH,76KuToDy.NaOH,96KaSuHi}, there is a lack of reliable rotation-vibration information for KOH and NaOH. Previous studies have tried to establish the fundamental vibrational wavenumbers but the situation is unclear with some debate in the literature and differing values presented (discussed further in Section~\ref{sec:spectra_koh_naoh}). Thus, the process of generating molecular line lists for KOH and NaOH is far more reliant on theory, particularly the quantum chemical methods used to build the spectroscopic model of the molecule, i.e.\ the potential energy and dipole moment surfaces. These first-principles methodologies~\citep{jt654,jt626} are routinely employed by the ExoMol project~\citep{jt528} and enable the computation of line lists containing many billions of molecular transitions \citep{jt631,jt810}. Importantly, these methods can enable the production of complete line lists suitable for elevated temperatures. Already a large number of diatomic and polyatomic species have been treated within the ExoMol framework~\citep{jt731}, with many line lists relevant to the spectroscopy of hot rocky super-Earths.

In this paper, we present comprehensive molecular line lists for KOH and NaOH that are suitable for modelling exoplanet and other hot atmospheres with temperatures up to 3500~K. The line lists have been generated using robust first-principles calculations based on newly computed, high-level \textit{ab initio} potential energy and dipole moment surfaces. Work similar in spirit was recently carried out on another hot rocky super-Earth molecule, silicon dioxide (SiO$_2$)~\citep{jt797}, resulting in a line list containing nearly 33 billion transitions between 5.69 million rotation-vibration states with rotational excitation up to $J=255$, where $J$ is the total angular momentum quantum number of the molecule.

\section{Methods}
\label{sec:methods}

A full description of the computational approach used to generate the molecular line lists of KOH and NaOH is provided in the supplementary material and only a brief summary is given here. The methodology is very similar to our previous work on silicon dioxide~\citep{jt797}. To begin with, high-level electronic structure methods were employed to construct the potential energy surface (PES) and electric dipole moment surface (DMS) of each molecule in its ground electronic state. Particular emphasis was placed on the accuracy of the \textit{ab initio} methods used to compute the PESs, as there is very little experimental data on these molecules and the PESs would not be empirically refined. Generally speaking, with the chosen \textit{ab initio} methods, the PES of KOH can be expected to reproduce the fundamental term values to within 3--5~cm$^{-1}$, while the PES of NaOH to within 1--3~cm$^{-1}$ as conservative estimates (e.g.\ see \citet{19OwYaKu.CH3F} and references within). Transition intensities computed using \textit{ab initio} DMSs are comparable to, if not occasionally more reliable, than experiment~\citep{13Yurchenko.method,jt573}, and for KOH and NaOH we expect them to be within 5--10\% of experimentally determined intensities.

All surfaces were computed on extensive grids of nuclear geometries (tens of thousands of points) with energies up to $h c \cdot 30\,000$~cm$^{-1}$ ($h$ is the Planck constant and $c$ is the speed of light). The \textit{ab initio} data was then fitted with appropriate three-dimensional analytic representations and incorporated into variational calculations with the nuclear motion code \textsc{TROVE}~\citep{TROVE}. Line list calculations for KOH and NaOH were very similar and followed standard \textsc{TROVE} procedures, notably taking advantage of a new linear triatomic molecule implementation~\citep{20YuMexx} that was recently used to compute high-temperature line lists for silicon dioxide~\citep{jt797} and carbon dioxide~\citep{jt804}. Rovibrational calculations utilised a large symmetry-adapted basis set with testing to ensure convergence of the $J=0$ vibrational states. Note that the potential energy and dipole moment surfaces are provided as supplementary material along with Fortran routines to construct them.

The KOH OYT4 line list was computed with a lower state energy threshold of $h c \cdot 16\,000$~cm$^{-1}$ ($h$ is the Planck constant and $c$ is the speed of light) and considered transitions up to $J=255$ in the 0\,--\,6000~cm$^{-1}$ range. A total of 38\,362\,078\,911 transitions between 7\,307\,923 energy levels were computed for the OYT4 line list. For the NaOH OYT5 line list, a lower state energy threshold of $h c \cdot 16\,000$~cm$^{-1}$ was used with transitions considered up to $J=206$ in the 0\,--\,9000~cm$^{-1}$ range. A total of 49\,663\,923\,0921 transitions between 7\,927\,877 energy levels were computed for the OYT5 line list. Figures showing the distribution of transitions and energy levels of each line list as a function of $J$ are shown in the supplementary material. In both molecules, the largest number of computed transitions and states occurs before $J=30$ and then steadily decreases as a result of an upper state energy threshold in the line list calculations.

Commenting on the quantum numbers and molecular symmetry of KOH and NaOH, the rovibrational energy levels and wavefunctions of both molecules were classified under the $\bm{C}_{\mathrm{s}}\mathrm{(M)}$ molecular symmetry group~\citep{98BuJexx} where the nuclear spin statistical weights are $g_{\mathrm{ns}}=\lbrace 8,8\rbrace$ for states of symmetry $\lbrace A^{\prime},A^{\prime\prime}\rbrace$, respectively. Transitions follow the symmetry selection rules $A^{\prime} \leftrightarrow A^{\prime\prime}$; and the standard rotational selection rules, $J\p-J\pp=0,\pm 1,\; J\p+J\pp \ne 0$; where $\p$ and $\pp$ denote the upper and lower state, respectively. Traditionally, the vibrational and rotational modes of asymmetric linear molecules are classified according to irreducible representations of $\bm{C}_{\mathrm{\infty v}}\mathrm{(M)}$: $\Sigma^+$/$\Sigma^{-}$, $\Pi$, $\Delta$ etc. The overall rovibrational wavefunctions can only be of $\Sigma^{+}$ or $\Sigma^{-}$ symmetry (see, e.g. \citet{98BuJexx}), which correlate to $\bm{C}_{\mathrm{s}}\mathrm{(M)}$ as  $A^{\prime} \leftrightarrow \Sigma^{+}$ and $A^{\prime\prime} \leftrightarrow \Sigma^{-}$. Another standard spectroscopic descriptor is the Kronig parity $e/f$~\citep{75BrHoHu.diatom}, related to the total parity $+1$ ($A^{\prime}$) and  $-1$ ($A^{\prime\prime}$) as follows: the parity of the $e$ state is $(-1)^J$ while the parity of the $f$ state is $(-1)^{J+1}$. The vibrational quantum numbers used by \textsc{TROVE} (discussed further in the supplementary material) were correlated to the following standard spectroscopic quantum numbers used for linear-type triatomic molecules: $v_1$, $v_2^{\rm lin}$, $L=|l|$, $v_3$, where $v_1$ and $v_3$ are the stretching quantum numbers, $v_2^{\rm lin}$ is the bending quantum number used for linear molecules and $l$ is the corresponding vibrational quantum number. The two bending quantum numbers $v_2^{\rm lin}$ and $l$ are related to the `non-linear' bending quantum number $v_2$ by $v_2^{\rm lin} = 2 v_2 + l$ with $L = v_2^{\rm lin}, v_2^{\rm lin} - 2, \ldots 0 (1)$. The symmetries of the vibrational and rotational contributions span the $A^{\prime}$ and $A^{\prime\prime}$ irreducible representations in
$\bm{C}_{\mathrm{s}}\mathrm{(M)}$ and $\Sigma^{+/-}$ ($L=0$), $\Pi$ ($L=1$), $\Delta$ ($L=2$) etc. in $\bm{C}_{\mathrm{\infty v}}$. The rotational quantum number $k_a$ is constrained to the vibrational angular momentum by $k_a=l$.

\section{Results}
\label{sec:results}

\subsection{Line list format}

As standard, the OYT4 and OYT5 line lists are provided in the ExoMol data format. The most up-to-date information on the ExoMol data format can be found in the recent ExoMol 2020 database release paper~\citep{jt810}. Further information can also be found in the ExoMol 2016 database release paper~\citep{jt631} if needed. In short, the \texttt{.states} file, see Table~\ref{tab:states} for an example of the KOH OYT4 \texttt{.states} file (identical structure for the NaOH OYT5 line list), contains all the computed rovibrational energy levels (in cm$^{-1}$), each labelled with a unique state ID counting number, symmetry and quantum number labelling, and the contribution $C_i$ from the largest eigencoefficient used to assign the rovibrational state. The \texttt{.trans} files contain all the computed transitions with upper and lower state ID labels and Einstein $A$ coefficients, see Table~\ref{tab:trans} for an example of the KOH OYT4 \texttt{.trans} file (identical structure for the NaOH OYT5 line list), and have been separated into $100$~cm$^{-1}$ frequency bins for user-handling purposes. In addition to the complete OYT4 and OYT5 line lists, which contain tens of billions of lines, more compact versions of these line lists, e.g.\ using the super-lines format, will be made available on the ExoMol website (\href{http://www.exomol.com}{www.exomol.com}) in due course.

\begin{table*}
\caption{\label{tab:states}Extract from the \texttt{.states} file of the KOH OYT4 line list.}
\begin{threeparttable}
{\tt
\centering
\tabcolsep=5pt
\begin{tabular}{rrccrccccccccc}
\hline\hline\\[-3mm]
\multicolumn{1}{r}{$i$} & \multicolumn{1}{c}{$\tilde{E}$} & $g_{\rm tot}$ & $J$ & unc & $\Gamma_{\rm tot}$& $v_1$ & $v_2^{\rm lin}$ & $L$ &  $v_3$ &$C_i$ & $n_1$ & $n_2$ & $n_3$\\
\hline\\[-3mm]
1	&	0.000000	&	8	&	0	&	4.000000	&	$A^{\prime}$	&	0	&	0	&	0	&	0	&	1.00	&	0	&	0	&	0	\\
2	&	441.302845	&	8	&	0	&	4.000000	&	$A^{\prime}$	&	1	&	0	&	0	&	0	&	1.00	&	1	&	0	&	0	\\
3	&	655.924321	&	8	&	0	&	8.000000	&	$A^{\prime}$	&	0	&	2	&	0	&	0	&	1.00	&	0	&	0	&	1	\\
4	&	877.113398	&	8	&	0	&	8.000000	&	$A^{\prime}$	&	2	&	0	&	0	&	0	&	1.00	&	2	&	0	&	0	\\
5	&	1081.543317	&	8	&	0	&	12.000000	&	$A^{\prime}$	&	1	&	2	&	0	&	0	&	1.00	&	1	&	0	&	1	\\
6	&	1269.577195	&	8	&	0	&	16.000000	&	$A^{\prime}$	&	0	&	4	&	0	&	0	&	1.00	&	0	&	0	&	2	\\
7	&	1307.451689	&	8	&	0	&	12.000000	&	$A^{\prime}$	&	3	&	0	&	0	&	0	&	1.00	&	3	&	0	&	0	\\
8	&	1501.902919	&	8	&	0	&	16.000000	&	$A^{\prime}$	&	2	&	2	&	0	&	0	&	1.00	&	2	&	0	&	1	\\
9	&	1687.220972	&	8	&	0	&	20.000000	&	$A^{\prime}$	&	1	&	4	&	0	&	0	&	1.00	&	1	&	0	&	2	\\
10	&	1732.331589	&	8	&	0	&	16.000000	&	$A^{\prime}$	&	4	&	0	&	0	&	0	&	1.00	&	4	&	0	&	0	\\
\hline\hline
\end{tabular}
}
\begin{tablenotes}
\item $i$: State counting number;
\item $\tilde{E}$: Term value (in cm$^{-1}$);
\item $g_{\rm tot}$: Total state degeneracy;
\item $J$: Total angular momentum quantum number;
\item unc: Estimated uncertainty of energy level (in cm$^{-1}$);
\item $\Gamma_{\rm tot}$: Overall symmetry in $\bm{C}_{\mathrm{s}}\mathrm{(M)}$ ($A^{\prime}$ or $A^{\prime\prime}$);
\item $v_1$, $v_2^{\rm lin}$, $L$, $v_3$: Linear-molecule vibrational quantum numbers;
\item $C_i$: Largest coefficient used in the \textsc{TROVE} assignment;
\item $n_1$, $n_2$, $n_3$: \textsc{TROVE} vibrational quantum numbers.
\end{tablenotes}
\end{threeparttable}
\end{table*}

\begin{table}
\centering
{\tt
\tabcolsep=10pt
\caption{\label{tab:trans} Extract from the \texttt{.trans} file for the $0$\,--\,$100\,$cm$^{-1}$ window of the KOH OYT4 line list.}
\begin{tabular}{rrr}
\hline\hline\\[-3mm]
\multicolumn{1}{c}{$f$}  &  \multicolumn{1}{c}{$i$} & \multicolumn{1}{c}{$A_{if}$}\\
\hline\\[-3mm]
2997983	&	3068742	&	5.87023848E-05	\\
656704	&	579657	&	2.18505406E-02	\\
832462	&	859993	&	7.59010951E-06	\\
1583619	&	1610495	&	1.26488417E-04	\\
243548	&	225006	&	7.77221741E-10	\\
3760055	&	3692577	&	4.66414104E-03	\\
4615110	&	4673426	&	1.49445735E-05	\\
582879	&	607777	&	1.20942018E-04	\\
5072485	&	5090674	&	2.24346842E-05	\\
5391819	&	5376327	&	4.14086508E-05	\\
\hline\hline\\[-2mm]
\end{tabular}
}
\\
\noindent
\footnotesize{
$f$: Upper  state ID; $i$:  Lower state ID; \\
$A_{if}$:  Einstein $A$ coefficient (in s$^{-1}$).
}
\end{table}

\subsection{Temperature-dependent partition functions}
\label{sec:pfn}

\begin{figure}
\centering
\includegraphics[width=0.48\textwidth]{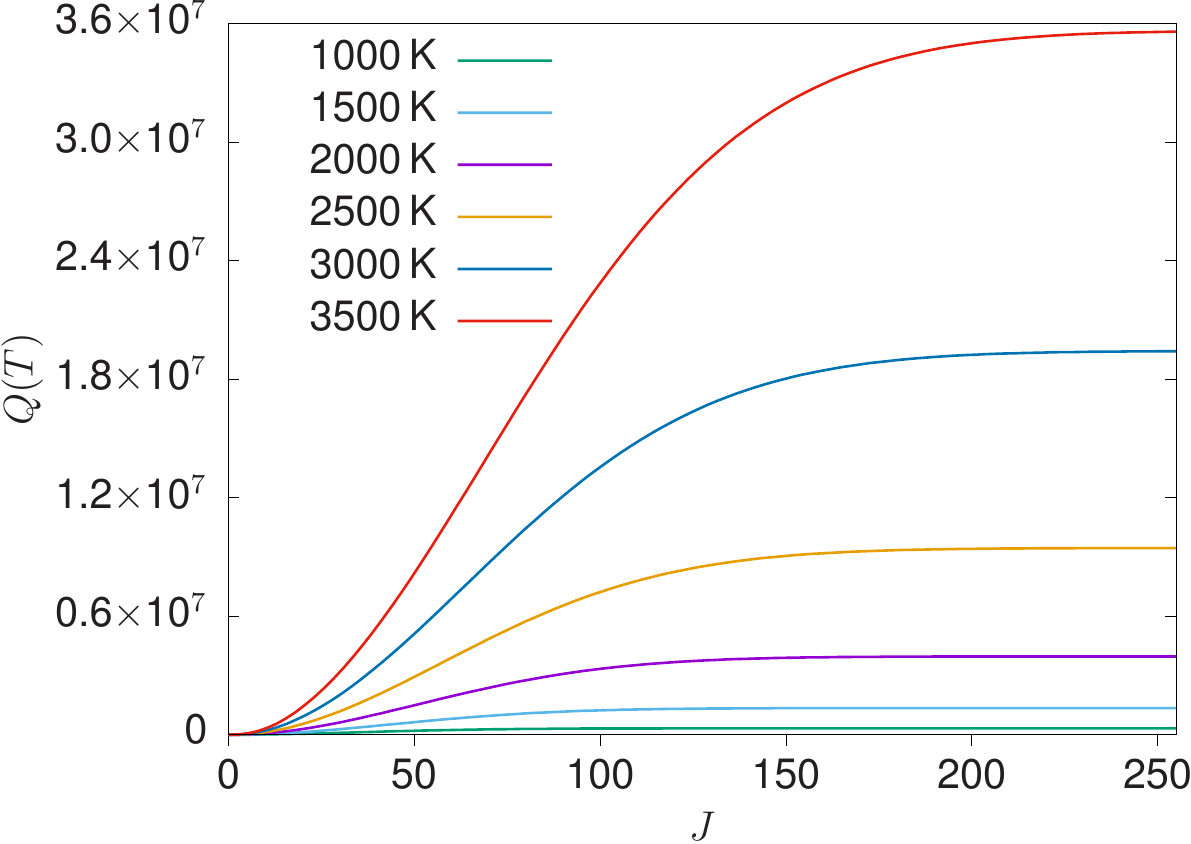}
\includegraphics[width=0.48\textwidth]{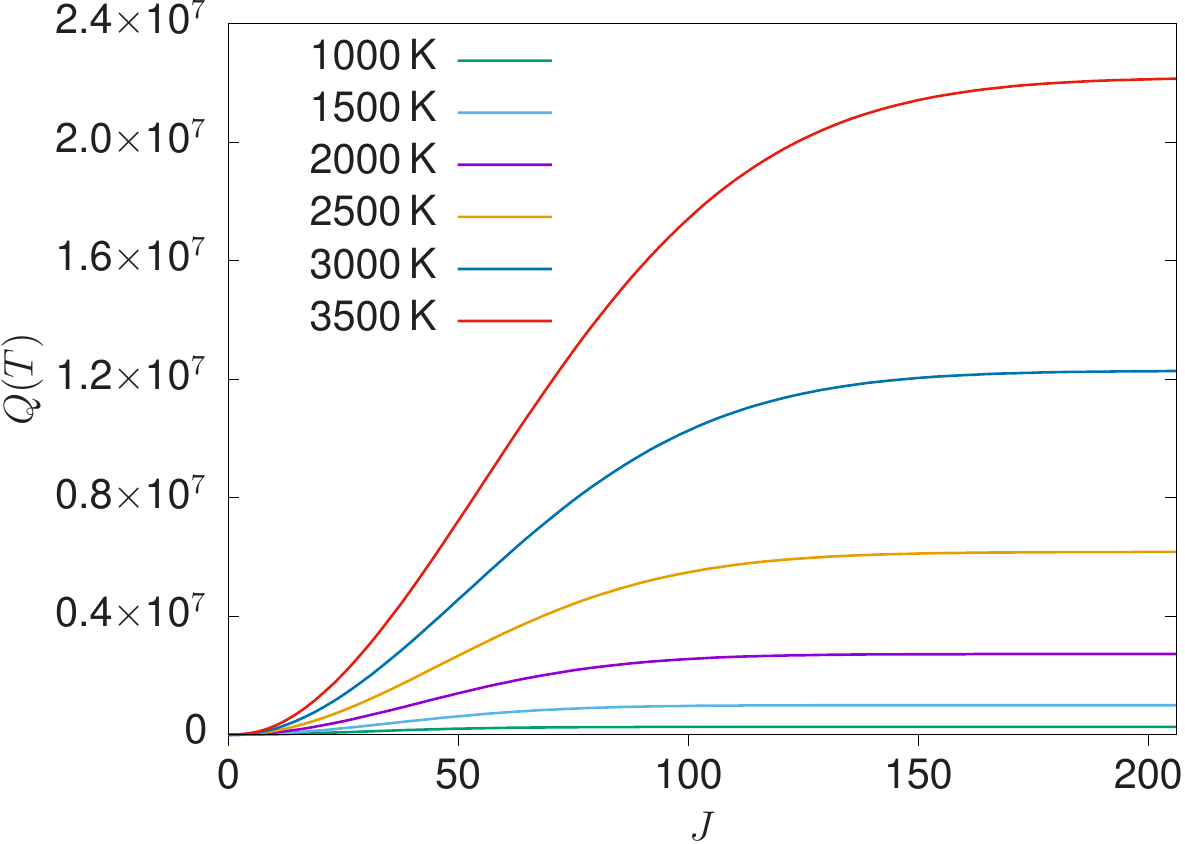}
\caption{\label{fig:pfn}Convergence of the partition function $Q(T)$ of KOH (left panel) and NaOH (right panel) with respect to the rotational quantum number $J$ at different temperatures.}
\end{figure}

The temperature-dependent partition function $Q(T)$, defined as
\begin{equation}
\label{eq:pfn}
Q(T)=\sum_{i} g_i \exp\left(\frac{-E_i}{kT}\right) ,
\end{equation}
where $g_i=g_{\rm ns}(2J_i+1)$ is the degeneracy of a state $i$ with energy $E_i$ and rotational quantum number $J_i$, has been calculated for KOH and NaOH by summing over all computed rovibrational energy levels on a $1$~K grid in the 1\,--\,4000~K range (provided as supplementary material). The convergence of $Q(T)$ as a function of $J$ for different temperatures is shown in Fig.~\ref{fig:pfn}. As expected, the partition function converges quickly at lower temperatures but this changes dramatically for temperatures above 2500~K, where a significant number of higher $J$ states must be considered in the summation of Eq.~\eqref{eq:pfn}. At $J=255$, the value of $Q(3500\,{\rm K})$ is converged to 0.007\% for KOH, while for NaOH the value of $Q(3500\,{\rm K})$ at $J=206$ is converged to 0.012\%.

\begin{figure}
\centering
\includegraphics[width=0.48\textwidth]{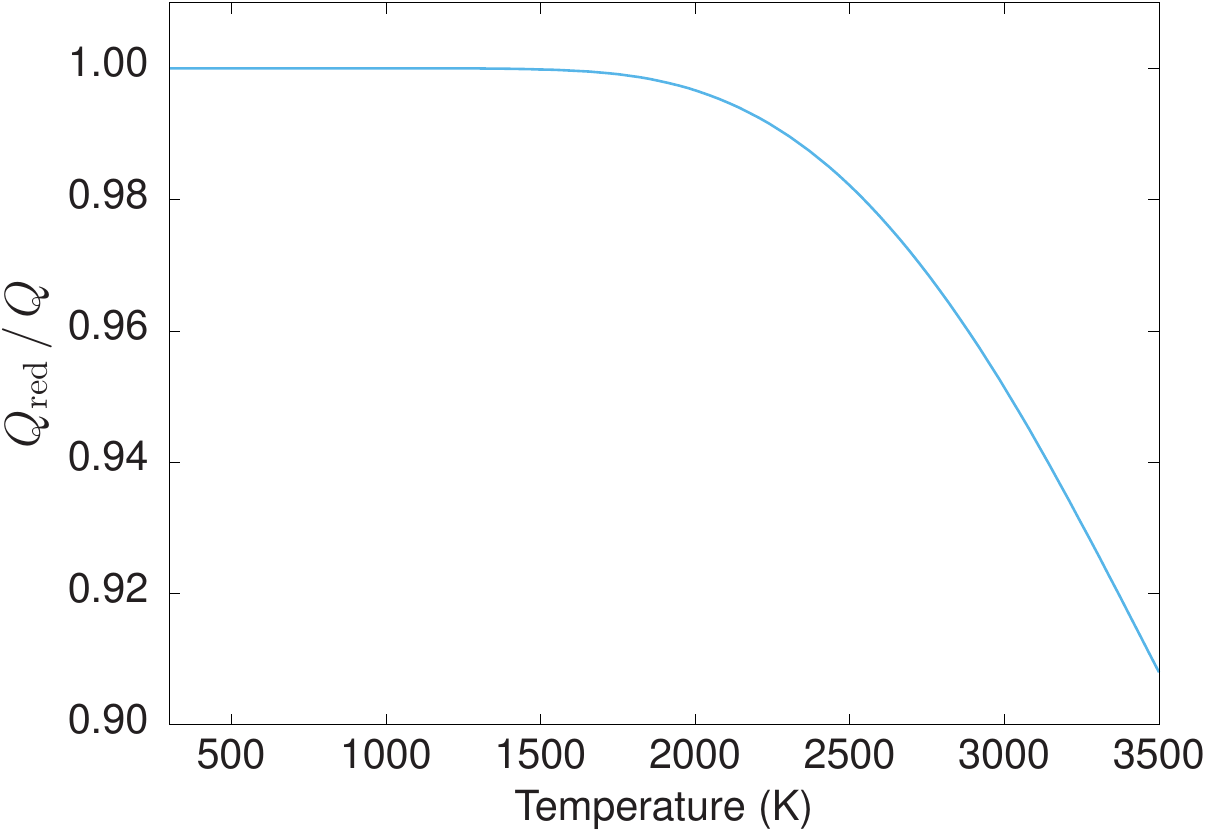}
\includegraphics[width=0.48\textwidth]{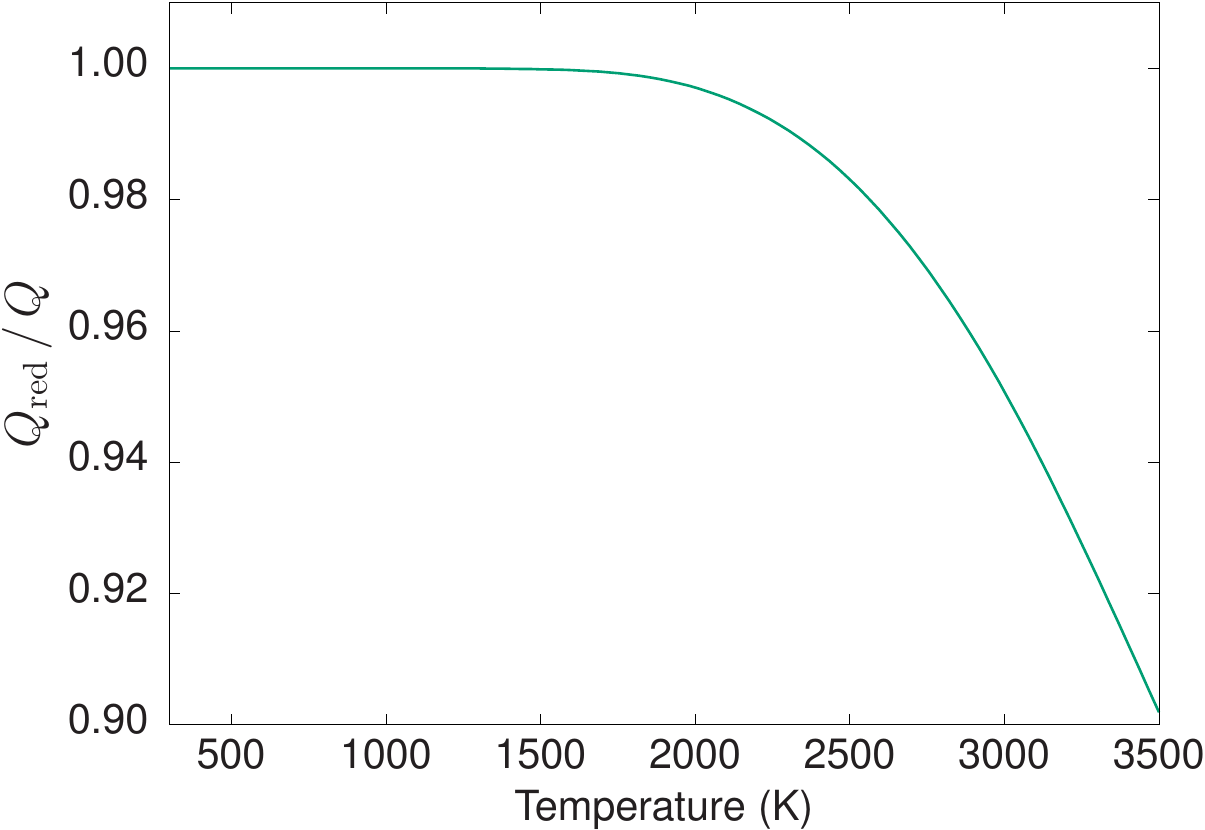}
\caption{\label{fig:pfn_lim}The ratio $Q_{\rm red}/Q$ as a function of temperature $T$ for KOH (left panel) and NaOH (right panel); this gives a measure of the completeness of the line lists.}
\end{figure}

Both the KOH and NaOH line lists were computed with a lower state energy threshold of $h c \cdot 16\,000$~cm$^{-1}$. To gain insight into the completeness of these line lists it is informative to calculate the reduced partition function $Q_{\rm red}(T)$, which only includes energy levels up to $h c \cdot 16\,000$~cm$^{-1}$ in the summation of Eq.~\eqref{eq:pfn}. In Fig.~\ref{fig:pfn_lim}, the ratio $Q_{\rm red}(T)/Q(T)$ has been plotted with respect to temperature for KOH (left panel) and NaOH (right panel). At lower temperatures $Q_{\rm red}/Q=1.00$, indicating completeness of the line list as the partition function values are not altered by the lower state energy threshold. Above 1500~K the ratio starts to decrease from unity and by 3500~K the ratio $Q_{\rm red}/Q\approx0.90$. This should be regarded as a soft temperature limit for the OYT4 and OYT5 line lists. Any use of these line lists above this temperature may result in a progressive loss of opacity, however, the missing opacity contribution can be estimated from $Q_{\rm red}/Q$ if needed~\citep{jt181}.

\subsection{Simulated spectra of KOH and NaOH}
\label{sec:spectra_koh_naoh}

\begin{figure}
\centering
\includegraphics[width=0.7\textwidth]{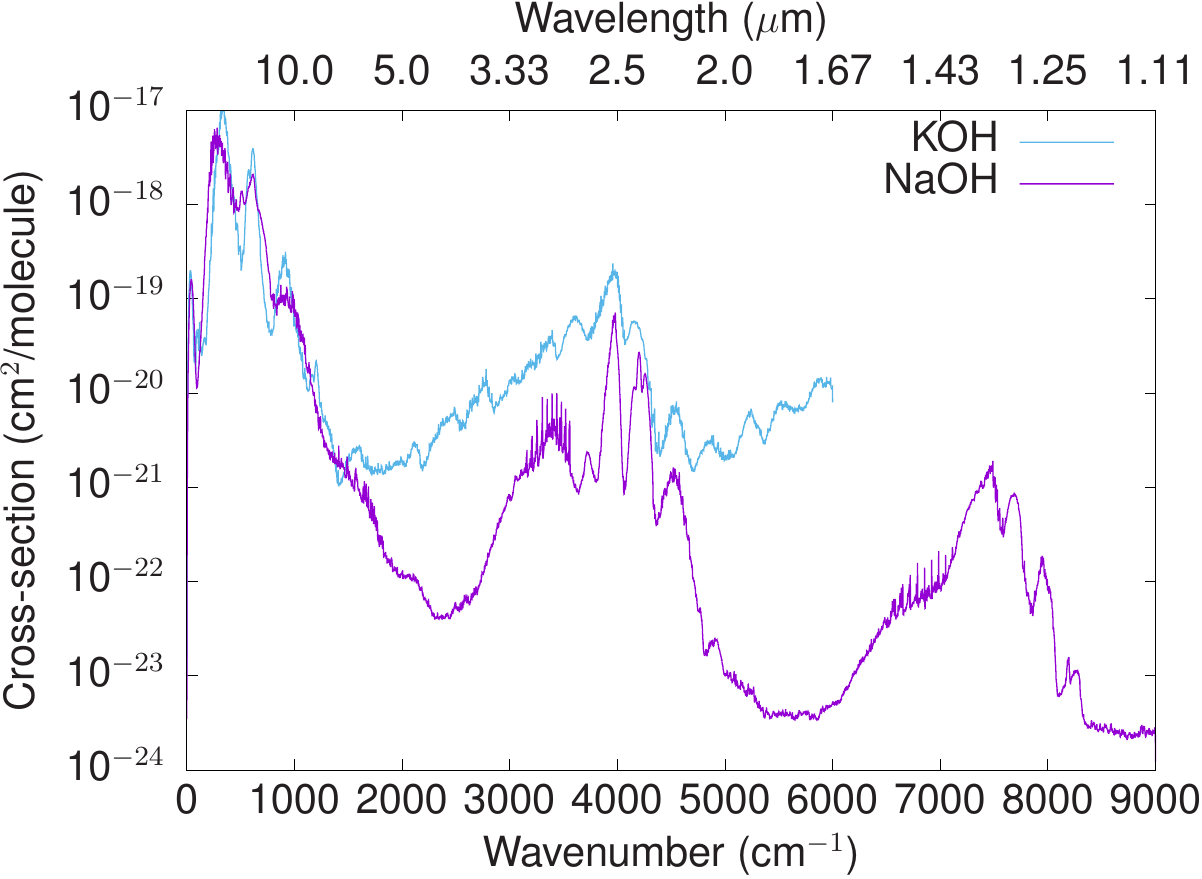}
\caption{\label{fig:1000K_both}Spectra of KOH and NaOH at $T=1000$~K. Absorption cross-sections were computed from the OYT4 and OYT5 line lists at a resolution of 1~cm$^{-1}$ and modelled with a Gaussian line profile with a half width at half maximum (HWHM) of 1~cm$^{-1}$.}
\end{figure}

An overview of the spectrum of KOH and NaOH is shown in Fig.~\ref{fig:1000K_both}, where we have simulated integrated absorption cross-sections at a resolution of 1~cm$^{-1}$, modelled with a Gaussian line profile with a half width at half maximum (HWHM) of 1~cm$^{-1}$. Spectral simulations were carried out with the \textsc{ExoCross} program~\citep{jt708}. The line positions of the strongest absorption bands are in close agreement between KOH and NaOH, which is to be expected given their structural similarity. Both molecules possess strong fundamental bands below 1000~cm$^{-1}$ (wavelengths $\lambda > 10$~$\mu$m) attributed to the $\nu_2$ bending mode at $343$~cm$^{-1}$ in KOH and $228$~cm$^{-1}$ in NaOH, and the $\nu_1$ stretching mode around $441$~cm$^{-1}$ in KOH and $562$~cm$^{-1}$ in NaOH. The next strongest band in both molecules is due to the fundamental $\nu_3$ O--H stretching mode, occurring around $3735$~cm$^{-1}$ in KOH and $3784$~cm$^{-1}$ in NaOH.

The values given above for the fundamental wavenumbers have been calculated in this study. In the past, efforts have been made to establish the values of the fundamentals of KOH and NaOH but the situation remains unclear and there has been some debate in the literature with differing values presented. The reader is referred to the section on molecular constants on pages 1046--1050 of \citet{97GuBeGo.KOH} for KOH and pages 1261--1263 of \citet{96GuBeGo.NaOH} for NaOH. These two papers provide a description of previous spectroscopic studies on these molecules and then go on to estimate values for their own work on the calculation of thermal functions. For KOH, these values are: $\nu_1=408\pm10$~cm$^{-1}$, $\nu_2=300\pm10$~cm$^{-1}$, $\nu_3=3700\pm100$~cm$^{-1}$; and for NaOH: $\nu_1=540\pm20$~cm$^{-1}$, $\nu_2=300\pm20$~cm$^{-1}$, $\nu_3=3700\pm100$~cm$^{-1}$. Despite not being in complete agreement with these estimated values, we are confident that our calculated fundamental wavenumbers are in fact closer to the true values based on the \textit{ab initio} theory used in this work to construct the spectroscopic models. For KOH, the fundamental bands should be reliable to within 3--5~cm$^{-1}$ and for NaOH to within 1--3~cm$^{-1}$ as conservative estimates. We have performed further checks by computing harmonic vibrational frequencies using high-level \textit{ab initio} methods with the quantum chemistry program \textsc{Molpro}~\citep{MOLPRO} and the results are consistent with our predictions.

\begin{figure}
\centering
\includegraphics[width=0.48\textwidth]{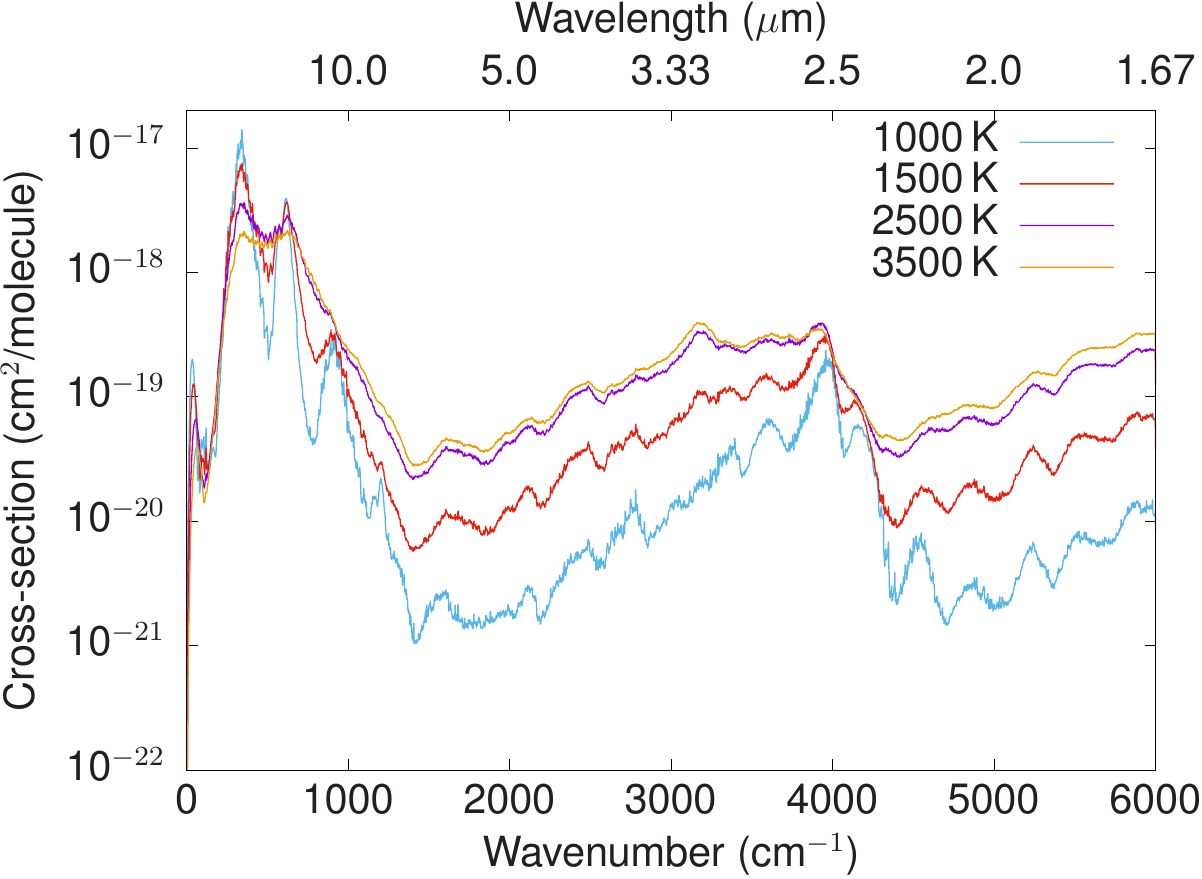}
\includegraphics[width=0.48\textwidth]{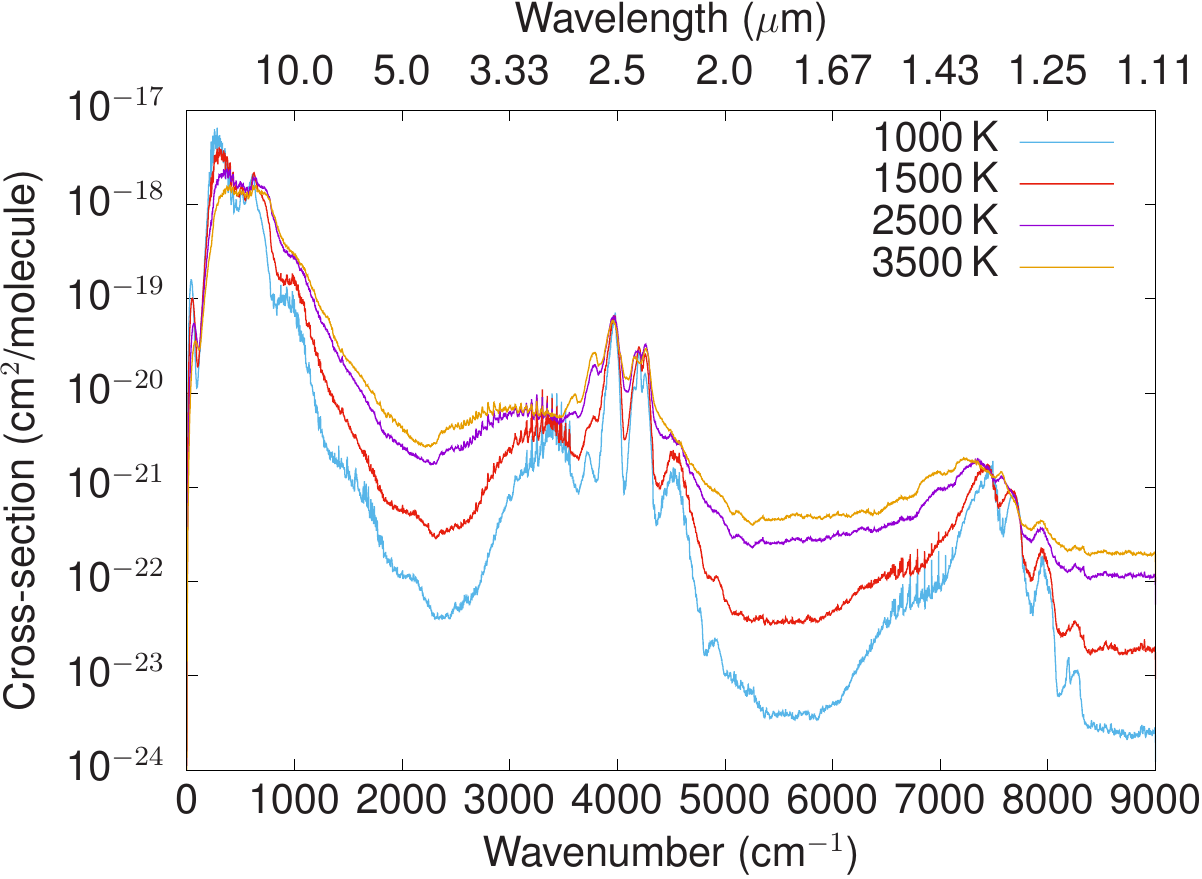}
\caption{\label{fig:temp_koh_naoh}Temperature dependence of the spectrum of KOH (left panel) and NaOH (right panel). Absorption cross-sections were computed from the OYT4 and OYT5 line lists  at a resolution of 1~cm$^{-1}$ and modelled with a Gaussian line profile with a half width at half maximum (HWHM) of 1~cm$^{-1}$.}
\end{figure}

The temperature dependence of the OYT4 and OYT5 line lists is shown in Fig.~\ref{fig:temp_koh_naoh}, where absorption cross-sections at a resolution of $1$~cm$^{-1}$ using a Gaussian profile with a HWHM of $1$~cm$^{-1}$ have been plotted for a variety of temperatures. As the temperature increases the rotational band envelopes are significantly broadened by the increased population of vibrationally excited states and the spectrum becomes smoother and more featureless, particularly in KOH above 1000~cm$^{-1}$ (wavelengths $\lambda < 10$~$\mu$m).

The Cologne Database for Molecular Spectroscopy (CDMS)~\citep{CDMS:2001,CDMS:2005} contains a number of pure rotational transitions in the microwave region for these molecules. For KOH, CDMS lists 61 lines up to $J=61$, determined using an effective Hamiltonian model based on observed microwave transitions from \citet{76PeWiTr.KOH,75KuToDy.KOH,87RaYaGia.KOH,96KaSuHi}. The CDMS KOH transition intensities are based on an experimentally-determined electric dipole moment value of 7.415~Debye in the ground vibrational state~\citep{96CeOlRi.KOH}. Regarding NaOH, there are 72 lines up to $J=72$ based on measurements from \citet{73PeTrxx.NaOH,76KuToDy.NaOH,96KaSuHi} with line intensities derived from a dipole moment value of 6.832~Debye~\citep{96KaSuHi}. Note that we have compared with the hyperfine-unresolved data from CDMS.

\begin{figure}
\centering
\includegraphics[width=0.48\textwidth]{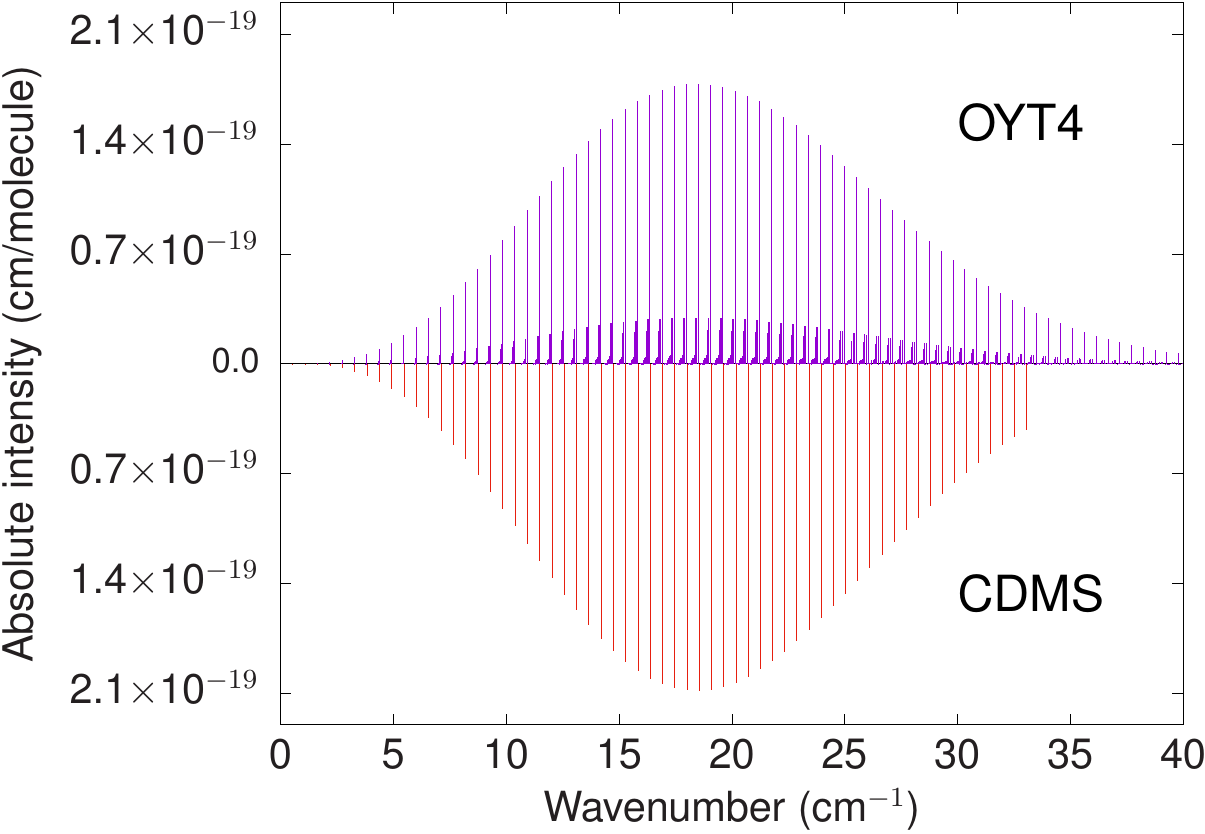}
\includegraphics[width=0.48\textwidth]{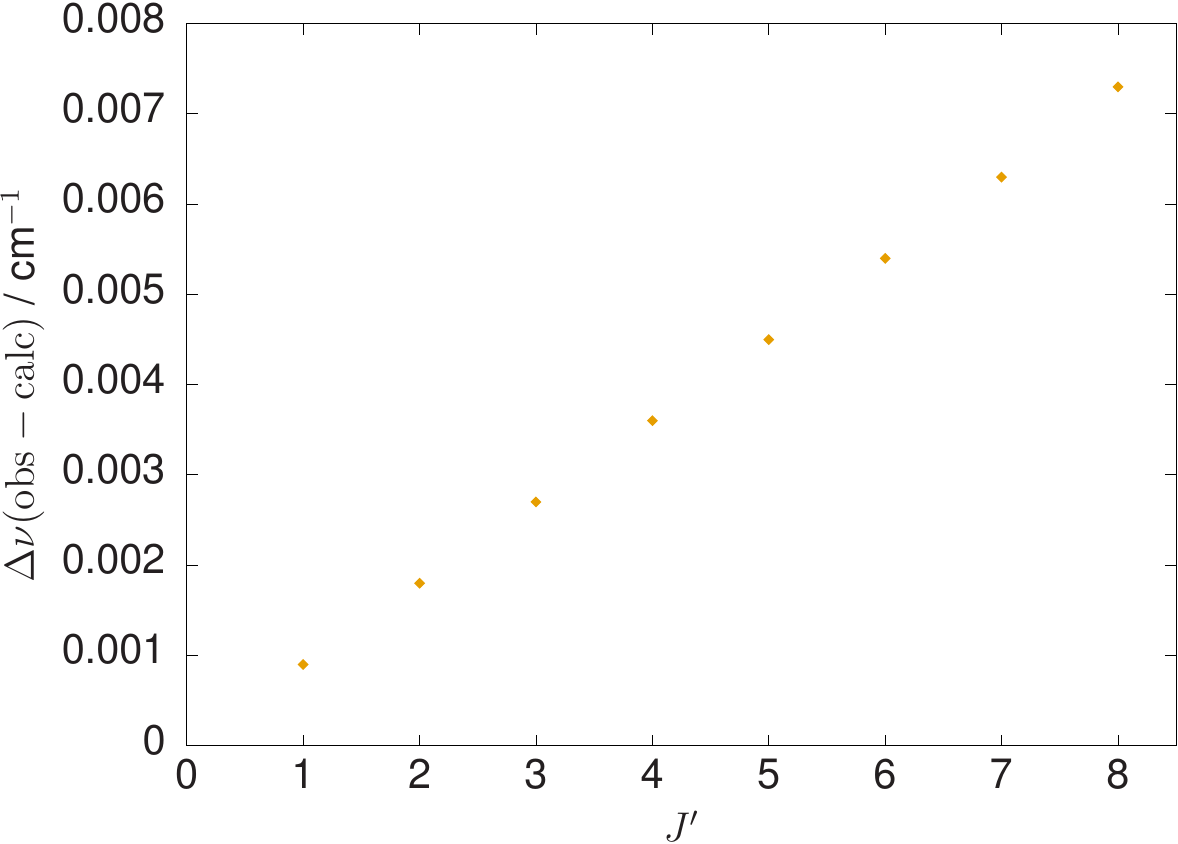}
\caption{\label{fig:rot_koh}Comparison of the KOH OYT4 line list with CDMS~\citep{CDMS:2001,CDMS:2005}. The left panel shows an OYT4 stick spectrum at $T=300$~K plotted against all transition data from CDMS. The right panel shows the residual errors $\Delta\nu({\rm obs}-{\rm calc})$ (in cm$^{-1}$) between the observed CDMS wavenumbers and the computed OYT4 wavenumbers up to $J^{\prime}=8$, where $J^{\prime}$ is the rotational quantum number of the upper state involved in the pure rotational transition.}
\end{figure}

In Fig.~\ref{fig:rot_koh}, the KOH OYT4 stick spectrum at $T=300$~K is plotted against the data from CDMS (left panel). The rotational band is reproduced well, with the OYT4 spectrum exhibiting more structure and lines as it extends up to $J=255$. The right panel of Fig.~\ref{fig:rot_koh} shows the residual errors $\Delta\nu({\rm obs}-{\rm calc})$ (in cm$^{-1}$) between the observed CDMS wavenumbers and the computed OYT4 wavenumbers up to $J^{\prime}=8$, where $J^{\prime}$ is the rotational quantum number of the upper state involved in the pure rotational transition, e.g.\ $J^{\prime}=1$ corresponds to the $1\leftarrow 0$ line. At each step up in $J^{\prime}$, the residual error increases by $\approx 0.001$~cm$^{-1}$. This is a very small error and confirms the quality of our computational approach, notably the accuracy of the \textit{ab initio} PES. Computed OYT4 line intensities are marginally smaller compared to the CDMS values and we are inclined to believe that the OYT4 values are more reliable since they are based on a high-quality \textit{ab initio} DMS and more sophisticated model.

\begin{figure}
\centering
\includegraphics[width=0.48\textwidth]{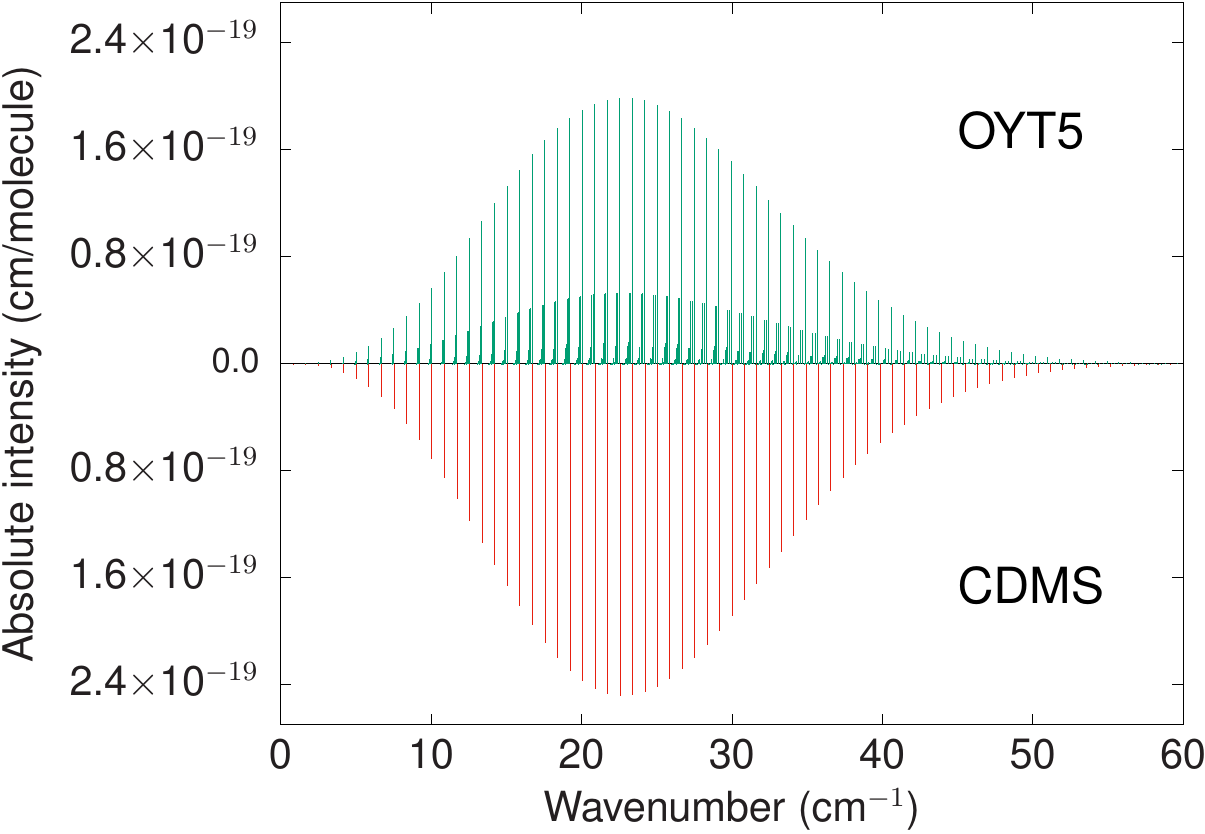}
\includegraphics[width=0.48\textwidth]{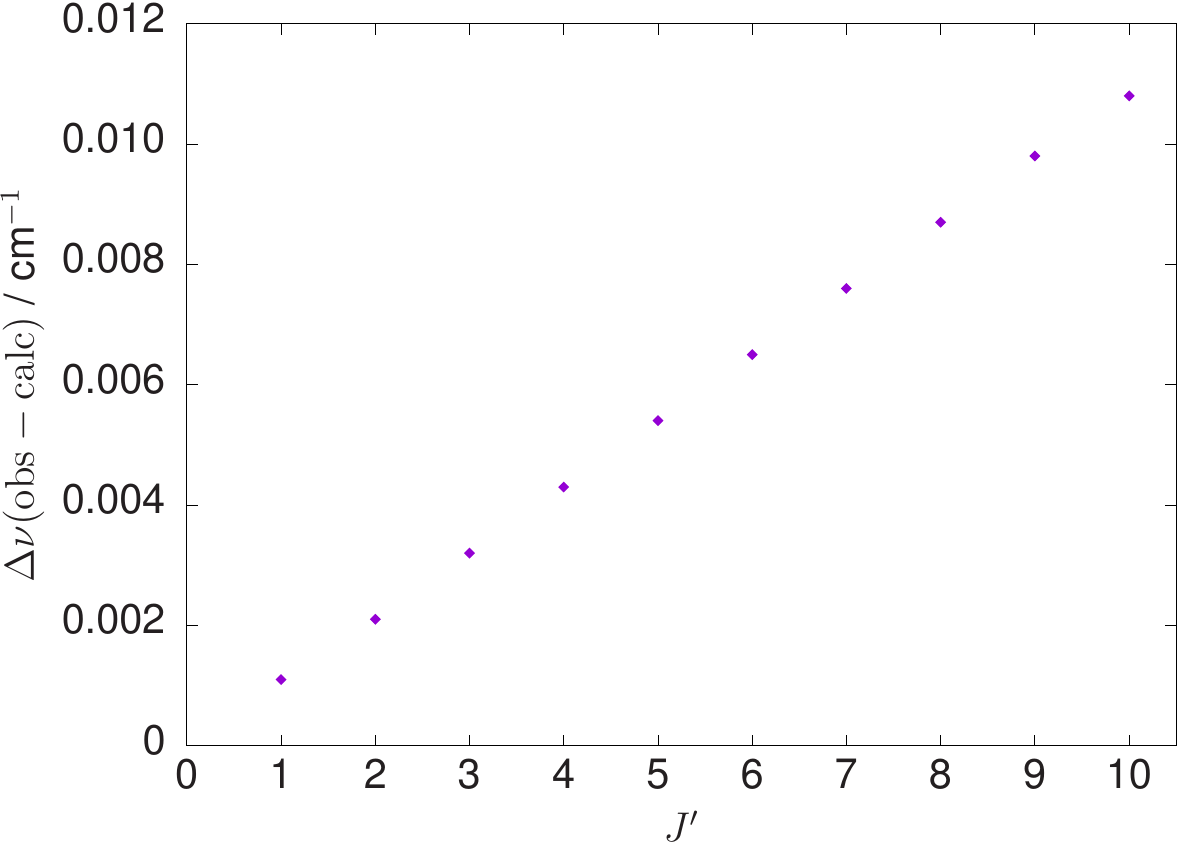}
\caption{\label{fig:rot_naoh}Comparison of the NaOH OYT5 line list with CDMS~\citep{CDMS:2001,CDMS:2005}. The left panel shows an OYT5 stick spectrum at $T=300$~K plotted against all transition data from CDMS. The right panel shows the residual errors $\Delta\nu({\rm obs}-{\rm calc})$ (in cm$^{-1}$) between the observed CDMS wavenumbers and the computed OYT5 wavenumbers up to $J^{\prime}=10$, where $J^{\prime}$ is the rotational quantum number of the upper state involved in the pure rotational transition.}
\end{figure}

A similar picture is obtained for the NaOH OYT5 line list when compared with the CDMS data, which can be seen in Fig.~\ref{fig:rot_naoh}. Rotational band shape is well reproduced with the OYT5 spectrum showing more structure due to the increased coverage up to $J=206$. Again the residual errors $\Delta\nu({\rm obs}-{\rm calc})$ are small and increase by $\approx 0.001$~cm$^{-1}$ for each $J^{\prime}$; a reflection of the quality of the \textit{ab initio} PES. The computed OYT5 line intensities are also weaker than the CDMS values but are still believed to be more reliable. Overall there is good agreement for both KOH and NaOH with the CDMS data.

\section{Conclusion}
\label{sec:conc}

Comprehensive molecular line lists have been presented for the alkali metal hydroxides KOH and NaOH. The KOH OYT4 line list covers the 0\,--\,6000~cm$^{-1}$ range (wavelengths $\lambda > 1.67$~$\mu$m) for states below $J=255$ and is applicable for temperatures up to 3500~K. The NaOH OYT5 line list covers the 0\,--\,9000~cm$^{-1}$ range (wavelengths $\lambda > 1.11$~$\mu$m) for states below $J=206$ and is also applicable for temperatures up to 3500~K. Both line lists have been constructed using purely \textit{ab initio} methods with no stage of empirical refinement to the spectroscopic model, thus limiting the accuracy of the predicted line positions, particularly for highly excited states and shorter wavelengths. For KOH, the fundamental bands should be reliable to within 3--5~cm$^{-1}$ and for NaOH to within 1--3~cm$^{-1}$ as conservative estimates. Computed transition intensities are expected to be within 5--10\% of experimentally determined intensities. These are, however, only error estimates based on our previous experience building \textit{ab initio} spectroscopic models with similar electronic structure methods. To truly quantify the accuracy of the presented line lists would require reliable experimental data and we encourage further spectroscopic investigations into these systems. Certainly the comparisons of the OYT4 and OYT5 pure rotational bands with the CDMS data are encouraging and give us confidence in our calculations.

Usually the ExoMol procedure to construct a line list will incorporate a stage of refinement of the spectroscopic model to laboratory data to improve the accuracy of the line list, notably the line positions~\citep{jt511}. Whilst this has not been possible for KOH and NaOH, purely \textit{ab initio} line lists can still be valuable in interpreting astronomical spectra. For example, the original hydrogen cyanide (HCN) / hydrogen isocyanide (HNC) line list of \citet{jt298} was computed at a lower level of \textit{ab initio} theory~\citep{jt273} than that used in the present study. This HCN / HNC line list has underpinned many recent (possible) detections of HCN in exoplanets~\citep{jt629,18HaMaCa.HCN,jt782}, having been updated~\citep{jt374,jt570}, and even adapted to H$^{13}$CN~\citep{jt447}, by the {\it post hoc} insertion of empirical energy levels, a process that is explicitly allowed for by the ExoMol data format~\citep{jt548}. Should high-resolution spectroscopic data become available for KOH and NaOH, a similar update will occur. With that in mind, the OYT4 and OYT5 line lists are only recommended for low-resolution studies of exoplanet atmospheres, e.g.\ with a resolving power of $R \approx 100$, and are not suitable for high-resolution analysis.

While the extraterrestrial presence of NaOH and KOH largely relies on the predictions of models, the spectrum of another hydroxide, CaOH, is thought to play an important role in the  atmospheres of M-dwarf stars \citep{72Pesch.CaOH,73Tsuji.CaOH,85BeBrxx.CaOH,06DiShWa.CaOH}. Indeed, \citet{13RaReAl.CaOH} studied missing opacity in the BT-Settl cool star model for M-dwarfs which they assigned to the lack of line lists for three species, namely AlH, NaH and CaOH.
The ExoMol project has since provided line lists for AlH \citep{jt732} and NaH \citep{jt605} leaving CaOH as the outstanding
problem. However, CaOH is a more complicated system than NaOH or KOH since it has an extra, unpaired electron meaning that it is an open
shell system with low-lying electronic states. It is these states which provide the CaOH spectral features and opacities in
M-dwarfs. While some work has been done on computing rovibronic transition intensities \citep{jt697}, so far we have not computed
a full rovibronic line list for a triatomic system. Work is currently underway adapting the program EVEREST~\citep{EVEREST},
which is designed to treat the Renner-Teller effect in the excited electronic states found in CaOH, for this purpose. A full rovibronic
line list for CaOH will be published as part of the ExoMol series in due course.

\section*{Acknowledgments}

This work was supported by the STFC Projects No. ST/M001334/1 and ST/R000476/1. The authors acknowledge the use of the UCL Legion High Performance Computing Facility (Legion@UCL) and associated support services in the completion of this work, along with the Cambridge Service for Data Driven Discovery (CSD3), part of which is operated by the University of Cambridge Research Computing on behalf of the STFC DiRAC HPC Facility (www.dirac.ac.uk). The DiRAC component of CSD3 was funded by BEIS capital funding via STFC capital grants ST/P002307/1 and ST/R002452/1 and STFC operations grant ST/R00689X/1. DiRAC is part of the National e-Infrastructure.

\section*{Data Availability}

The data underlying this article are available from the ExoMol database at \href{http://www.exomol.com}{www.exomol.com}.


\section*{Supporting Information}
Supplementary data are available at MNRAS online. This includes a detailed description of the methodology, the potential energy and dipole moment surfaces of KOH and NaOH with programs to construct them, and values of the temperature-dependent partition function of KOH and NaOH up to 4000~K. The following references were cited in the supplementary material: \citep{17HiPexx.ai,05LiScMe.ai,02PeDuxx.ai,92KeDuHa.ai,MOLPRO,Molpro:JCP:2020,01TyTaSc.H2S,97PaScxx.H2O,03Watson.methods,93JoJexx.H2O,jt797,98CsAlSc,07AdKnWe.ai,08PeAdWe.ai,10HiPexx.ai,09HiPeKn.ai,04TenNo.ai,08YoPexx.ai,02Weigend.ai,05Hattig.ai,11YaYuRi.H2CS,10HiMaPe.ai,05KaGaxx.ai,08KaGaxx.ai,MRCC:JCP:2020,CFOUR:JCP:2020,89Dunning.ai,74DoKrxx.ai,86Hess.ai,01deHaDi.ai,TROVE,jt466,15YaYuxx.method,jt626,
17YuYaOv.methods,20YuMexx,83CaHaSu,jt96,24Numerov.method,61Cooley.method,jt46,jt708}

\label{lastpage}

\end{document}